# Comment

**Dennis Lindley**

I was taught by Harold Jeffreys, having attended his postgraduate lectures at Cambridge in the academic year 1946–1947, and also knew him when I joined the Faculty there. I thought I appreciated the *Theory of Probability* rather well, so was astonished to read this splendid paper, which so successfully sheds new light on the book by placing it in the context of recent developments.

Jeffreys's—he preferred that form of the possessive—main aim in writing the Theory, his term for TP, was to provide tools for scientists, like himself, more famous then for his geophysics, to use in the observational data they encountered. In the preface to the second edition, he criticizes reviewers of the first for the fact that "no mention was made of the fact that the book contained useful methods of treatment of several problems of practical importance." It is primarily a text on operational statistics. This is most strikingly seen in his development of significance tests, producing results that are distinct from those of Fisher, who was also at Cambridge, though the distinction was not apparent to either of them then. Cambridge was then, as it still is, a true university in the sense that you would regularly meet people outside your own, often narrow, discipline, in college activities. In this atmosphere, Jeffreys was much influenced by a group of philosophers including W. E. Johnson, C. D. Broad and J. M. Keynes, and, as a result, thought seriously about the scientific method, where he was also influenced by Karl Pearson's *Grammar of Science*. (In my view, the best thing KP ever wrote.) It is this atmosphere of data collection in astronomy, combined with the philosophy of science, that produced the Theory; an atmosphere in which mathematics is an essential tool, but only a tool. His attitude to mathematics is best seen in the magisterial book he wrote with his wife, *Methods of Mathematical Physics* (Jeffreys and Swirles, 1946). In light of these considerations, it is clear that his respect for mathematical rigor, while high, did not occupy a dominant position; it was the application that mattered. Robert and his colleagues are right to criticize Jeffreys's attitude to improper distributions but, if uniform over the whole real line gave a sensible posterior, that was good enough for him. He did notice the difficulties with several variances.

There is one point in the Theory where, in my view, he makes an error that he might have recognized. It occurs in equation (1) in Section 3.10 when, in modern terms, he integrates over the sample space to produce the invariants needed for his objective priors. In retrospect, it is surprising that he did this, especially when, elsewhere in the Theory, he condemns the use of integration over the tails of distributions, so incorporating results that did not occur, in the common, non-Bayesian form of a significance test. As a result of the integration in equation (1) the invariant prior can depend on the experiment to be performed; that is, the sample space to be used. Thus the invariant prior for a chance $\theta$ would differ according to whether you were going to use direct, or inverse, binomial sampling. Chance $\theta$ was, for Jeffreys, a representation of a real thing and ignorance of it should not depend on how it was to be studied. I did not appreciate this issue until Birnbaum introduced me to the likelihood principle.

This error, in a sense, arises from a disputed philosophical view of the nature of science. Jeffreys, like many scientists, both then and now, regarded the scientific method as objective; indeed objectivity was held to be one of, if not the principal, advantages of science over other ways of understanding the world. It was his search for objectivity, in the form of a definition of ignorance, that led him to violate the likelihood principle, which he had recognized rather informally in the condemnation of tails mentioned above. It is obvious now, and should have been at the time of the first edition in 1946, that there are subjective elements in the scientific method as when, in the early stages of an investigation, scientists disagree because of the limited data available. It is only with the accumulation of more evidence that agreement is reached and apparent objectivity obtained. Statistical methods, as Haldane pointed out, are most valuable with modest amounts of data. Jeffreys's error left the way for de Finetti and Savage to lay the foundations for Bayesian ideas in a coherent way.

Let me turn from errors to his triumphs, and the great concepts that he introduced. One of these is his Chapter 1 in which he states, and produces a "proof" that uncertainties, always present with modest amounts of data, must obey the basic rules of probability. It is not, as some eclectic statisticians say today, that one has a choice; one does not, probability is the unique tool. Although he never refers to them in this context, he was effectively saying that Neyman and Pearson were wrong. Confidence intervals and tail-area significance levels, are not probability statements about the quantity of interest and therefore do not satisfy the requirements of his Chapter 1. Notice that Jeffreys *proved* that assertion about probability. The authors of this paper are correct to question the proof, for it does not even stand up to the mathematics of 1939, as we in the audience saw in 1947, but it makes an important first step. Actually Ramsey was ahead of Jeffreys, both in time and rigor, and it is astonishing the he did not know of Ramsey's work, for he lived literally just down the road. When, in the 1950s, I pointed this out to him, Jeffreys was also astonished, for he had been at Ramsey's death bed. What they had established was that one had to be a Bayesian, there was no logical choice.

In their perceptive analysis, the authors remind me that I must have learned from Jeffreys the fact, to which I now attach much importance, that probability is always a function of two arguments. It is a defect of much modern instruction in elementary statistics that this is unrecognized and we talk of the probability of an event without mentioning the conditions under which the uncertainty is being contemplated.

His second triumph was a general method for the construction of significance tests, putting a concentration of prior probability on the null value—no ignorance here—and evaluating the posterior probability using what we now call Bayes factors. He was not only disagreeing with Neyman and Pearson, but also with Popper, whose philosophy of science was, and regrettably still is, popular among scientists. Jeffreys told me that "Popper can't do probability," and that he had opposed Popper's election to the Royal Society. Bayesians take Jeffreys's method for granted because it can be used effectively in so many situations. His work on estimation is less striking and he was opposed to the use of a point estimate. The only estimate was the posterior density of the parameter being considered. His distinction between probability and chance (page 5) is valuable. Chance is a property of sequences, which de Finetti later termed exchangeable, so that if you believe a sequence has this property, then you accept chance and may have beliefs, that is, probabilities, about its value. The distinction avoids the difficulties when probabilities of probabilities are introduced.

Much modern statistical literature discusses problems in a decision framework; for example, referring to a decision to reject a null hypothesis. Yet despite this, there is little statistical literature on practical decision problems, using a loss, or utility, function representing reality. The Cambridge of the 30s, and perhaps even later, was concerned with knowledge and learning, feeling that applications were outside their ivory towers and best left to others. The Theory reflects this attitude and the occasional references to decisions are incidental. In modern terms, he was concerned with the probability of the quantity of interest, given the data; and not with decisions about that quantity, decisions that Ramsey, influenced by Keynes, so beautifully discussed. With both Ramsey and Keynes, King's College appears more practically oriented than Jeffreys's St. Johns.